\documentclass[twocolumn,
               showpacs,
               preprintnumbers,
               nofootinbib,
               prd,
               superscriptaddress,
               10pt,
               notitlepage,
               aps]{revtex4-2}
\usepackage{graphicx,amssymb,amsmath,amsthm,amsfonts,epsfig,mathtools}

\usepackage[utf8]{inputenc}
\usepackage{graphicx}
\usepackage{dcolumn}
\usepackage{bm}
\usepackage{amsmath}
\usepackage{mathrsfs}
\usepackage{color}
\usepackage{soul}
\usepackage[dvipsnames]{xcolor}
\usepackage{hyperref}
\hypersetup{colorlinks=true, citecolor=MidnightBlue,linkcolor=CornflowerBlue, urlcolor=CornflowerBlue, linktocpage=true}
\usepackage{xfrac}
\def\mphi{m_{\phi}}

\def\rt{\tilde{\rho}}

\def\tg{\tilde{g}}

\def\meff{m_{\textrm{eff}}}
\def\betagb{\beta_{\rm GB}}
\def\sigmacrit{\sigma_{\rm crit}(\betagb)}

\begin{document}

\title{Subtleties in constraining gravity theories with mass-radius data}

\author{Ekrem S. Demirbo\u{g}a}
\email{edemirboga17@ku.edu.tr}
\affiliation{Department of Physics, Ko\c{c} University,
Rumelifeneri Yolu, 34450 Sariyer, Istanbul, Turkey}

\author{Yakup Emre \c{S}ahin}
\email{yakup.sahin@boun.edu.tr}
\affiliation{Department of Physics, Bo\u{g}azi\c{c}i University, 34342, Istanbul, Turkey}

\author{Fethi M. Ramazano\u{g}lu}
\email{framazanoglu@ku.edu.tr}
\affiliation{Department of Physics, Ko\c{c} University,
Rumelifeneri Yolu, 34450 Sariyer, Istanbul, Turkey}

\date{\today}

\begin{abstract}
Simultaneous measurements of neutron star masses and radii can be used to constrain deviations from general relativity (GR) as was recently demonstrated for the spontaneous scalarization model of Damour and Esposito-Farèse (DEF). Here, we investigate the general applicability of the same procedure beyond this single example. We first show that a simple variation of the DEF model renders the same mass-radius measurements ineffective for obtaining constraints. On the other hand, a recently popular and distinct model of scalarization that arises in scalar-Gauss-Bonnet theory can be constrained similarly to the original DEF model, albeit due to a slightly different underlying mechanism. These establish that using the mass-radius data can potentially constrain various theories of gravity, but the method also has limitations.
\end{abstract}

\maketitle

\section{Introduction}
\label{sec:intro}
Recent advances in the observation of black holes and neutron stars have led to a wealth of information on both the astrophysics of these systems and the nature of gravity~\cite{Berti:2015itd,LIGOScientific:2018dkp,Barack:2018yly}. Gravitational waves have been the major novel player in this arena, but electromagnetic observations have also improved substantially, one example being the x-ray observations from neutron stars~\cite{Bogdanov:2019ixe,Bogdanov:2019qjb,Bogdanov:2021yip}.

Proper modeling of observational data can lead to the determination of the structure of a neutron star, one ingenious example being the simultaneous inference of a neutron star's mass and radius using x-ray emissions~\cite{Ozel:2015fia}. Once this data is obtained, it can be used to investigate the underlying physics. For example, for a given stellar mass, the radius of the star mainly depends on the equation of state (EOS) of nuclear matter. Hence, one can invert this relationship, and use the mass-radius data to determine the EOS, or more properly, constrain it in a parametrized form~\cite{Ozel:2015fia}.

A second factor that affects the mass-radius relationship of a neutron star is potential deviations from general relativity (GR). Most astrophysical studies assume GR in their analysis, as was the case for the determination of EOS we mentioned~\cite{Ozel:2015fia}, but the Tolman-Oppenheimer-Volkov (TOV) equations that are used to compute the stellar structure can change drastically if the governing theory of gravity deviates from that of Einstein~\cite{Eksi:2014wia}. Similarly to the case of EOS, we can use this relationship in the reverse direction and utilize the mass-radius data to constrain deviations from GR. Here, we will explore the feasibility of this idea.

Using the mass-radius data to constrain gravity theories has various challenges. We already mentioned that the nuclear EOS is also a determining factor of the stellar structure, but we do not know the exact EOS. Hence, there is already a fundamental level of uncertainty in the mass-radius relationship independently of how gravity behaves. Secondary factors such as the spin of the neutron star only make the error bars larger. Moreover, there are substantial assumptions in the modeling of the events that lead to the inference of mass and radius from the x-ray data, bringing yet more uncertainty~\cite{Ozel:2015fia,Tuna:2022qqr}. Overall, there is little hope to investigate small deviations from GR in the near future using mass-radius data. However, some of the most commonly studied alternative theories of gravity predict large observational signals, hence the relevance of our discussion.

Any theory of gravity which largely deviates from GR in all respects has already been ruled out, since we have already tested gravity in the weak-field regime of the Solar System~\cite{Will:2006LR}. Therefore, only theories that substantially change predictions for strong gravitational fields while closely following GR for the weak ones are relevant for our purposes. Scalar-tensor theories that feature \emph{spontaneous scalarization} phenomena are well-known examples of this~\cite{PhysRevLett.70.2220,Doneva:2022ewd}.

Many scalar-tensor theories admit GR solutions, but in spontaneous scalarization, these solutions can become unstable in the presence of a compact object, for example, a neutron star. Arbitrarily small scalar fields can grow exponentially around the star due to a tachyonic instability, but they eventually settle down to a stable configuration where the star is surrounded by a scalar cloud. A \emph{scalarized} star generically has a substantially different structure compared to its GR counterpart of the same mass, which is a hallmark of scalarization and one of its main appeals. Thus, theories featuring spontaneous scalarization, often simply called \emph{scalarization}, are ideal targets for investigation using mass-radius data. 

\textcite{Tuna:2022qqr} recently showed that spontaneous scalarization in the form originally conceived by Damour and Esposito-Farèse (DEF)~\cite{PhysRevLett.70.2220}, which is also the most commonly studied case, can indeed be constrained using the mass-radius data. One can obtain bounds on the theory parameters, or equivalently rule out some parts of the parameter space.

Even though \textcite{Tuna:2022qqr} demonstrated the power of mass-radius measurements in principle, it studied only a specific form of deviation from GR. Even when we restrict ourselves to scalarization, there are different mechanisms due to various coupling terms between the scalar fields and the metric, leading to very different observable signatures~\cite{Doneva:2022ewd}. Each choice of coupling leads to a distinct \emph{model} of scalarization, which is no less well-motivated or interesting than the original DEF model. Thus, it is  not known how widely applicable the mass-radius data is in constraining deviations from GR, even when we only consider theories that feature spontaneous scalarization. We will explore this topic by studying two widely popular scalarization models, and try to see how they differ from the original DEF model in terms of being constrained by the mass-radius data~\cite{Tuna:2022qqr}. 

The first example is an extension of the DEF model with a different coupling behavior at large scalar fields, $\phi \to \infty$.  This model is motivated by renormalization concerns and is also known to provide a radically different phenomenology when one of the main coupling constants of the theory, $\beta$, reverses its sign~\cite{PhysRevD.57.4802,Mendes:2014ufa,Palenzuela:2015ima,Mendes:2016fby,Mendes:2018qwo}. The second example features a coupling between the scalar and the Gauss-Bonnet term. Such \emph{scalar-Gauss-Bonnet} models are the most widely studied ones in recent years due to the fact that they allow scalarization of black holes in addition to neutron stars, unlike the original DEF model~\cite{Doneva:2017bvd,Silva:2017uqg}. 

We show that the mass-radius data is quite ineffective in constraining the first model. This is because the neutron star structures have relatively small deviations from GR compared to the original DEF model. We already mentioned that the mass-radius data is not expected to be very useful in such a situation, hence, this is also a lesson about spontaneous scalarization itself. Namely, some simple changes to the original DEF model can lead to theories where the promise of large deviations from GR is weakened, or even lost altogether. 

On the other hand, the mass-radius data is able to constrain the parameter space of the second model with Gauss-Bonnet coupling. The bounds we obtain are somewhat weak, but they demonstrate the effectiveness of our approach. Overall, we conclude that the mass-radius data is useful in testing alternative theories of gravity beyond the original DEF model, but it is by no means generically effective.  

{Our dataset here is slightly larger than that of \textcite{Tuna:2022qqr}, the previous study on this topic, and this led us to investigate how our analysis changes as the number of neutron star mass-radius measurements increases. A larger dataset leads to tighter bounds for the theory parameters, as expected. These improvements can be significant in some cases, and we observe that modest advances in observations can contribute meaningfully to our ability to test alternative theories of gravity.}

The rest of the paper is organized as follows. In Sec.~\ref{sec:scalarization} we explain the basic mechanism of scalarization and the specific scalarization models we study. In Sec.~\ref{sec:methods} we summarize the methods we use in order to obtain the mass-radius relationships theoretically predicted by the said models, and how the theory is compared to the observational data using Bayesian analysis. In Sec.~\ref{sec:results} we present the results of our analysis, {and in Sec.~\ref{sec:dataset_size} we discuss how they depend on the amount of data we use.} Finally in Sec.~\ref{sec:conclusions} we give our conclusions together with potential future directions. We use the gravitational units $G=1=c$.

\section{Scalarization in various models}
\label{sec:scalarization}
Here, we provide a summary of spontaneous scalarization and its extensions. A comprehensive review of the topic can be found in \textcite{Doneva:2022ewd}.

\subsection{Basics of spontaneous scalarization and the DEF model}
\label{sec:def_model}
The idea of spontaneous scalarization was first introduced by Damour and Esposito-Farèse for a scalar-tensor theory with the action~\cite{PhysRevLett.70.2220}
\begin{align}\label{st_action}
  \frac{1}{16\pi} &\int d^4x \sqrt{-g}\ \bigg[R
 -2g^{\mu\nu}  \nabla_{\mu} \phi  \nabla_{\nu} \phi\
  \bigg] \nonumber \\
 &+ S_{\rm m} \left[f_\text{m}, A^2(\phi) g_{\mu \nu} \right] .
\end{align}
This formulation is in the so-called Einstein frame where the gravitational part of the action is the same as GR. However, in the matter action $S_{\rm m}$, any matter field, denoted by $f_\text{m}$, couples to the conformally scaled metric $\tg_{\mu\nu} = A^2(\phi) g_{\mu \nu}$, the so-called Jordan frame metric.\footnote{Another formulation of the theory using the Jordan frame metric would have minimal matter coupling, but instead directly couple the Ricci scalar $\tilde{R}$ to the scalar field~\cite{Doneva:2022ewd}.} We will denote quantities such as the stress-energy tensor associated with this metric with tildes throughout.

In spontaneous scalarization, the uniform $\phi=0$ solutions correspond to GR, and they are valid solutions of the scalar-tensor theory. However, this trivial configuration becomes unstable around neutron stars, and any small scalar perturbation grows to reach another solution where the star is surrounded by a typically large scalar cloud~\cite{Doneva:2022ewd}. These clouds lead to potentially large deviations from GR near neutron stars, whereas there is no scalarization near less compact objects, meaning the weak field tests such as those within the Solar System are automatically passed.

The underlying mechanism for scalarization can be seen in the scalar equation of motion
\begin{align} \label{scalar_eom}
  \Box_g \phi  &= - 8 \pi  A^4 \frac{d\left( \ln A \right)}{d(\phi^2)}  \tilde{T}\ \phi \nonumber \\
  &\approx - 8 \pi \left[ A^4 \frac{d\left( \ln A \right)}{d(\phi^2)} \right]_{\phi=0} \tilde{T} \phi \equiv \meff^2\ \phi
\end{align}
where $\tilde{T}$ is the trace of the stress-energy tensor associated with $\tg_{\mu\nu}$, and we linearized the equation for small values of $\phi$ on the second line. Then, this becomes a generalized form of the massive wave equation where $\meff^2$ is a spacetime-dependent \emph{effective mass squared}. The key point is that $\meff^2$ can have either sign. When it is negative, small scalar field perturbations can behave as tachyons, leading to exponential growth. This growth can eventually stop due to nonlinear effects, and the scalar field can obtain a stable configuration. In other words, we say the star is \emph{scalarized}~\cite{Ramazanoglu:2016kul,Doneva:2022ewd}.

For a ``typical'' neutron star, the pressure is much smaller than the energy density, $\tilde{p} \ll \tilde{\rho}$. This means $\tilde{T}= -\tilde{\rho}+3\tilde{p} \approx -\tilde{\rho} <0$, where we assumed that the nuclear matter behaves as a perfect fluid. Thus, if the conformal coupling has a Taylor expansion of the form
\begin{align}\label{A_taylor}
   A(\phi) = 1+ \beta \phi^2/2 + \mathcal{O}(h^3),
\end{align}
with $\beta<0$, spontaneous scalarization can occur due to $\mphi^2$ becoming negative. This description might suggest that any matter distribution leads to scalarization when $\beta<0$, which is not the case. Outside the star, there is vacuum, hence $\mphi^2=0$. When this and other factors from spherical symmetry are taken into account, one can see that only sufficiently negative values of $\mphi^2$ lead to a tachyonic instability. In other words, for a given $\beta<0$, only stars with sufficiently large densities scalarize. For $|\beta| \sim 1$, this only occurs for neutron stars~\cite{Doneva:2022ewd,Ramazanoglu:2016kul}.

There are infinitely many choices for $A(\phi)$ that follows the pattern in Eq.~\eqref{A_taylor}, but by far the most commonly studied case is what we will call the \emph{quadratic DEF model}\footnote{Damour and Esposito-Farèse considered a few different functions $A(\phi)$ in their original study~\cite {PhysRevLett.70.2220}. However, Eq.~\eqref{eq:A_quadratic} became by far the most widely studied one in the subsequent literature, and has been sometimes simply called the \emph{DEF model}. We add the ``quadratic'' adjective to our naming to avoid confusion with other forms of $A(\phi)$.}
\begin{align}\label{eq:A_quadratic}
   A &= e^{\beta \phi^2/2} \ \ \ \  \textrm{\bf (quadratic DEF model)}
      \\
    &= 1+\frac{\beta}{2} \phi^2 +  \frac{\beta^2}{8} \phi^4 + \dots\ .
\end{align}
Our naming is due to the fact that in the $\phi \to \infty$ limit, the coupling function behaves as an exponential of a quadratic polynomial of the scalar: $\ln A(\phi \to \infty) \sim \phi^2$. Recall that what stops the tachyonic growth and provides the final stable scalarized solution is the nonlinear effects, hence the behavior of $A(\phi)$ in this asymptotic limit has an important role in determining the structure of the star.

So far, we only discussed the typical case of $\tilde{T}<0$ that requires $\beta<0$, however, $\tilde{T}$ can change sign near the core of massive enough neutron stars for some nuclear equations of state~\cite{PhysRevD.57.4802,Mendes:2014ufa,Palenzuela:2015ima,Mendes:2016fby,Mendes:2018qwo}. When $\tilde{T}>0$, the requirement for the onset of scalarization, $\mphi^2<0$, is satisfied only if $\beta>0$. Thinking in the reverse direction, for $\beta>0$, most neutron stars would not scalarize, but there could be exceptions when the stellar mass is high. This case is intimately connected to extensions of the DEF model beyond the quadratic case in Eq.~\eqref{eq:A_quadratic} as we will see next.

\subsection{Linear DEF model}
\label{sec:def_model}
There are various ways to modify the DEF model we have discussed so far. In this study, we will concentrate on how the form of $A(\phi)$ affects the results. The second form of $A(\phi)$ we will study is what we call the \emph{linear DEF model}
\begin{align}\label{eq:A_linear}
   A &= \left( \cosh \sqrt{3}\beta \phi \right)^{1/(3\beta)} \ \ \ \  \textrm{\bf (linear DEF model)}
   \\
    &= 1+\frac{1}{2} \beta \phi^2 +  \frac{\beta^2-2\beta^3}{8} \phi^4 + \dots\ ,
\end{align}
which was originally popularized by \textcite{Mendes:2016fby}. The naming is again due to the asymptotic behavior $\ln A(\phi \to \infty) \sim \phi$. Hence, the coupling changes considerably more slowly in the $\phi \to \infty$ limit compared to the quadratic DEF model, and consequently, the final scalarized solutions can be radically different from those of the quadratic DEF model.
 
What is the reason for the specific choice in Eq.~\eqref{eq:A_linear}? One motivation is related to the renormalizability of scalar-tensor theories~\cite{Mendes:2016fby}, but the main interest in the scalarization literature is due to the stability issues of scalarized solutions. 

We mentioned that the tachyonic instability that starts the scalarization process \emph{can} be quenched by nonlinear terms we ignore in the linearized equation~\eqref{scalar_eom}, but this is not always the case. A key finding is that all scalarized stars in the quadratic DEF model with $\beta>0$ are known to be unstable~\cite{PhysRevD.57.4802,Mendes:2014ufa,Palenzuela:2015ima,Mendes:2016fby,Mendes:2018qwo}. On the other hand, the linear DEF model of Eq.~\eqref{eq:A_linear} can have stable scalarized stars for $\beta>0$ as well, which means it is the only choice for this part of the parameter space out of the two models we have.

Our discussion of how the existence of a tachyonic instability depends on $\beta$ in the quadratic DEF model also holds for the linear DEF model, such as the appearance of scalarization only at sufficiently large values of stellar mass. However, unlike the quadratic DEF model that has been constrained using the mass-radius data~\cite{Tuna:2022qqr}, there has been no study on the linear DEF model. More generally, even though it is one of the more widely studied cases, the linear DEF model is known in much less detail compared to the quadratic one.

\subsection{Scalar-Gauss-Bonnet model}
\label{sec:sgb_model}
Spontaneous scalarization is not restricted to actions of the type in Eq.~\eqref{st_action}. Another scalar-tensor theory subfamily that provides a similar phenomenology is scalar Gauss-Bonnet (sGB) theories~\cite{Doneva:2017bvd,Silva:2017uqg,Doneva:2017duq}
\begin{align}\label{eq:sgb_action}
    S &= \frac{1}{16 \pi}\int d^4x \left[ R - 2 g^{\mu\nu}\nabla_{\nu} \phi \nabla_{\mu} \phi + \frac{\beta_{\rm GB}}{R_0^2} f(\phi) \mathscr{G} \right]
\end{align}
where
\begin{equation} \label{eq:def_gauss_bonnet}
\mathscr{G} = R^{\mu\nu\rho\sigma}R_{\mu\nu\rho\sigma}
- 4 R^{\mu\nu} R_{\mu\nu}
+ R^2,
\end{equation}
is the Gauss-Bonnet invariant, $f(\phi)$ is a coupling function and $\betagb$ is a dimensionless coupling constant. $R_0=1.47664$km is the typical length scale of the problem corresponding to one solar mass in geometric units, which is introduced to render $\betagb$ dimensionless~\cite{Doneva:2017duq}.

The scalar obeys
\begin{equation}\label{eq:sgb_eom}
    \Box \varphi =  -\beta_{\rm GB} \frac{d f}{d\phi} \mathscr{G} \approx -\beta_{\rm GB} \left. \frac{d^2 f}{d\phi^2} \right|_{\phi=0} \mathscr{G} \phi = \meff^2\ \phi,
\end{equation}
where we again linearized the equation around $\phi=0$ similarly to the DEF model in Eq.~\eqref{scalar_eom}. Hence, there are coupling functions $f(\phi)$ which lead to $\mphi^2<0$ and spontaneous scalarization. 

There are again infinitely many choices of $f(\phi)$ that allow scalarization. However, following the literature, we will concentrate on the single parameter $f(\phi)$ family~\cite{Doneva:2017bvd}\footnote{\textcite{Doneva:2017bvd} use different symbols: $\betagb \to \lambda^2$, $\sigma \to \beta$. We made these notation changes in order to avoid confusion with the parameters of the DEF models, and highlight the analogy of $\beta$ in the DEF models and $\betagb$ in the sGB model.}
\begin{align}\label{eq:f_parametrized}
    \betagb f(\phi) &= \betagb \frac{e^{-\sigma \phi^2}-1}{2\sigma} \ \ \ \  \textrm{\bf (sGB model)} \\
    &= -\frac{\betagb}{2} \phi^2 +  \frac{\betagb\sigma}{2} \phi^4 + \dots\ ,
\end{align}
where we explicitly keep the $\betagb$ factor to see the overall behavior of the coupling term. The asymptotic behavior $\phi \to \infty$ is similar to that of the quadratic DEF model, and we call this specific case the \emph{scalar-Gauss Bonnet (sGB) model}.

Note that the existence of the tachyon at the linear level is completely controlled by $\betagb$, $\meff^2 = \beta_{\rm GB} \mathscr{G} + \mathcal{O}(\phi^3)$, which is analogous to $\beta$ of the DEF models. $\betagb$ also controls the overall level of deviations from GR to a large degree, as we will discuss later. $\sigma$ does not play a role in the onset of scalarization, however, it controls the leading nonlinear corrections and the $\phi \to \infty$ regime, hence it significantly affects the mass and radius of a scalarized star. There is no analog of $\sigma$ in the DEF models which have the single parameter $\beta$ that controls the fully nonlinear as well as the linearized behavior of the scalar coupling.

\section{Methods}
\label{sec:methods}
%
\begin{table}
\begin{tabular}{|c || c | } 
 \hline
Model & Coupling \\ [0.5ex] 
 \hline\hline
Quadratic DEF & $A = e^{\beta \phi^2/2}\  \hspace{18mm} $ in Eq.~\eqref{st_action} \\
Linear DEF & $A = \left( \cosh \sqrt{3}\beta \phi \right)^{1/(3\beta)} \ \ $ in Eq.~\eqref{st_action}  \\
sGB & $f = \left[e^{-\sigma \phi^2}-1\right]/(2\sigma) \ \ \ $ in Eq.~\eqref{eq:sgb_action}  \\ [1ex]
 \hline
\end{tabular}
\caption{The three spontaneous scalarization models we discuss in this study.}
\label{table_model}
\end{table}
We want to compare the mass-radius diagrams implied by a given spontaneous scalarization model to observational data, so that we can obtain bounds on the model parameters. This comparison requires the following:
\begin{enumerate}
\item The computation of single neutron star structures for different models and a range of parameters

\item Repeating the single star computations for a range of slightly differing stars in order to obtain a mass-radius curve for each parameter set

\item Assigning a likelihood to the theoretical mass-radius curves, hence to the corresponding parameters, by comparison to the mass-radius data using Bayesian analysis

\end{enumerate}
These steps were implemented for the quadratic DEF model (and its massive scalar extension) in \textcite{Tuna:2022qqr}, which can be consulted for details. Here, we provide a summary, and also elaborate on the differences from that work. The three models of scalarization we investigate can be found in Table.~\ref{table_model} for easy referencing.

\subsection{Computing the mass-radius curves}
\label{sec:mass_radius}
%
The structure of a single neutron star for a given set of theory parameters, $\beta$ or $(\betagb,\sigma)$, is obtained by solving a modified version of the TOV equations~\cite{PhysRevLett.70.2220,Ramazanoglu:2016kul}. We cast these equations as a boundary value problem, and solve them using relaxation, an approach introduced by \textcite{Rosca-Mead:2020bzt} and later extended by \textcite{Tuna:2022qqr}. For this same set of model parameters, e.g. fixed values of $(\betagb, \sigma)$ in the sGB model, this process is repeated many times to obtain a continuous group of neutron stars differing by their masses and radii, which tells us how the mass of a star depends on its radius, the so-called mass-radius relationship. See Fig.~\ref{fig:gr_mr} for some simple examples. There are various sources of uncertainty in obtaining the mass-radius curve for a given scalarization model~\cite{Tuna:2022qqr}, which have important consequences for our methodology.

First, the structure of a neutron star is strongly dependent on how nuclear matter behaves inside them, which is not known to a high precision. Hence, deviations from GR are only one of the factors that determine the mass-radius curves. Indeed, the mass-radius dataset we utilize here was originally used to constrain the EOS of neutron star matter under the assumption that gravity is governed by GR~\cite{Ozel:2015fia}. For this reason, we repeat each computation for a variety of EOS and report individual cases as well as the probabilities that are marginalized over these EOS. This provides a measure of the effect of scalarization notwithstanding the EOS, but there is an intrinsic uncertainty nevertheless. We should note that this is not a problem specific to our case, any method that tries to investigate deviations from GR using neutron stars generally has to address this issue.\footnote{There are tests that are relatively insensitive to EOS, which go under the umbrella term of \emph{universal relations}~\cite{Yagi:2016bkt}.}

The second source of uncertainty in our results is the simplifying assumptions in our computations in order to make them tractable. All our neutron star mass and radius values are obtained assuming a spherically symmetric spacetime, however, astrophysical neutron stars do not have this symmetry due to their rotation. Rotation is typically a subleading factor in scalarization aside from the extreme cases, but ignoring its effect brings in another layer of uncertainty~\cite{Tuna:2022qqr,Yazadjiev:2016pcb}. 

Lastly, the mass-radius dataset we use was inferred under the assumption that the gravitational effect of the neutron star mass is the same everywhere outside the star. This is the case in GR since the spacetime can be described by the Schwarzschild metric, which has a single mass parameter $M$. However, the spacetime is more complicated when the scalar-tensor theory of our models are considered, and the analog of the mass parameter becomes a function that depends on the radial distance from the star, $M(r)$. Then, what is meant when we say the mass of the star? Whenever we talk about the neutron star mass in this study, we will exclusively mean the mass of the star gravitationally deduced at asymptotically far regions, the Arnowitt-Deser-Misner (ADM) mass. \textcite{Tuna:2022qqr} showed that the inferred results from the Bayesian analysis can be affected by this choice, but not in a qualitative way. Yet, this should be recognized as a source of uncertainty.

The overall result of the above points is that even if we find that the posterior likelihood of a scalarization model is larger than that of GR at face value, the statistical significance of such a result is hard to make precise. Furthermore, the same level of likelihood increase can also be obtained by ``optimizing'' our choice of EOS due to the degeneracy in the dependence of the mass-radius relationship on the EOS and deviations from GR. Because of this, we will be conservative and refrain from claims of detecting deviations from GR. 

On the other hand, the situation is not symmetrical when we try to rule out certain parameter values of a scalarization model, which is possible. For example, \textcite{Tuna:2022qqr} showed that in the $\beta \ll -1$ case of the quadratic DEF model, the posterior likelihood decays exponentially as $e^{-c |\beta|}$ for all EOS. This is because the mass-radius curves of this model become radically different from those of GR for such $\beta$ values. Perhaps more importantly, the curves move toward the $M > 2M_\odot$ region whereas the observational data largely shows the stars to have $M \lesssim 2M_\odot$ as we will discuss further later. This means the overlap of the theoretical curve with the observations becomes tiny, even when all modeling uncertainties are taken into account. Hence, such values can be ruled out. 

The approach we outlined provides a lower bound of $\beta \gtrsim -15$ on the quadratic DEF model~\cite{Tuna:2022qqr}. This is weaker than other bounds obtained from binary star observations~\cite{2013Sci...340..448A,Freire:2012mg,Zhao:2022vig}, but it shows that the idea of testing GR with mass-radius data works in principle. Furthermore, the only known bound on the quadratic DEF model when the scalar is massive was also obtained from the neutron star mass-radius data: $\beta \gtrsim -20$ given $\mphi \lesssim 2\times 10^{-11}$eV, $\mphi$ being the mass of the scalar field.

\subsection{Stability of solutions}
\label{sec:stability}
%
\begin{figure}
    \includegraphics[width=\columnwidth]{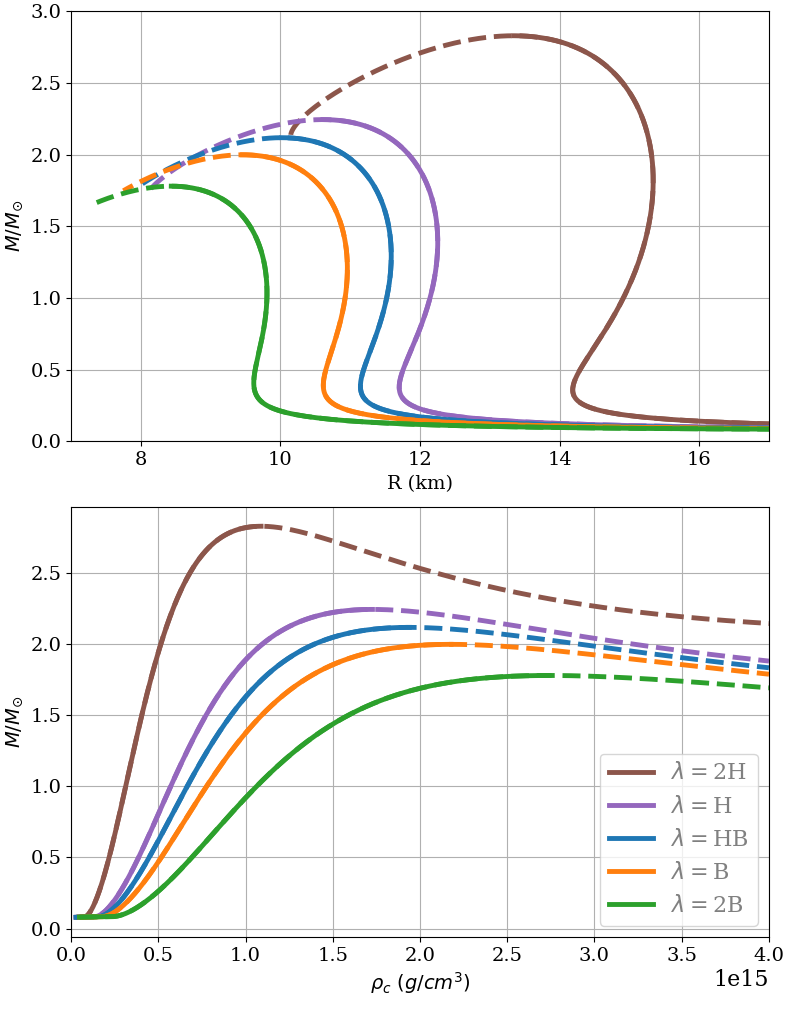}
    \caption{Hydrodynamically stable (solid lines) and unstable (dashed) portions of the mass-radius curves for different EOS under GR. See Sec.~\ref{sec:bayesian_analysis} for more information about the EOS.}
    \label{fig:gr_mr}
\end{figure}
An astrophysical object can never be completely isolated from its environment, hence, the only relevant solutions of the TOV equations, scalarized or not, are the ones that are stable under small perturbations. There are two main reasons a solution we compute can be unstable.

The first source of instability is hydrodynamic. How does the star behave if we perturb its matter distribution slightly, does it go back to the original configuration, or move away from it further? A commonly used, necessary but not sufficient, condition for hydrodynamic stability under GR is~\cite{Shapiro:1983du}
\begin{align}\label{eq:dmdrho}
\frac{dM}{d\rt_c}>0 \ ,
\end{align}
where $d\rt_c$ is the central density of the star. In a crude description, if a star does not obey this condition, it can decrease its total mass, hence energy, when its radius slightly decreases and its central density increases. Hence, inward radial perturbations are energetically preferred, so the star simply wants to collapse on itself. The situation is more complicated in scalar-tensor theories, but the same criterion is observed to hold for scalarized stars as well, which we will assume in all cases~\cite{Tuna:2022qqr,Mendes:2018qwo}. Hydrodynamically stable and unstable sections of the mass-radius curves under GR for various EOS can be seen in Fig.~\ref{fig:gr_mr}.

We should reiterate that condition~\eqref{eq:dmdrho} is not sufficient for hydrodynamical stability, and it is indeed known that there are neutron stars that satisfy this condition but are unstable. For us, the chief example is the $\beta>0$ case of the quadratic DEF model we mentioned. A detailed analysis of the quasinormal modes of scalarized stars in this model shows that some fluid perturbations grow exponentially rather than oscillate. Hence, there seems to be no astrophysically relevant scalarized solutions~\cite{Mendes:2018qwo}. Nevertheless, a short discussion of these stars can be found in Appendix~\ref{sec:appendix}.

The second source of instability is scalarization itself. We already mentioned that scalarization appears due to a tachyonic instability, which makes the scalar field grow starting from a GR solution. The growth can eventually stop due to nonlinear terms, leading to a stable scalarized neutron star, but this is not guaranteed. That is, there are cases where a neutron star of mass $M$ in GR is known to be unstable to scalarization, yet there is no stable scalarized solution at the same mass, or no scalarized solution at all. Then, what is the endpoint of the scalarization process? The star might lose some of its energy via gravitational and scalar radiation, moving to a less massive stable solution, or it might collapse on itself to end up as a black hole. Studies on the nonlinear evolution of such cases are few, but we know that there is no stable neutron star at such a mass, scalarized or not~\cite{Doneva:2022ewd}. This will be important for constructing the ultimate mass-radius curve for a given model.

How do we know that a star is tachyonically unstable? One can investigate any given star by linearizing the equations as in Eq.~\eqref{scalar_eom}, and then performing a spherical mode analysis to obtain Schr\"odinger-like equations~\cite{Doneva:2017bvd,Doneva:2022ewd}. However, the dependence of scalarization on the neutron star mass for each scalarization model makes the determination of stability easier. 

First, let us reexamine the quadratic DEF model with $\beta<0$, the most commonly studied scalarization scenario. We already mentioned that the tachyonic instability occurs only above a minimal stellar mass $M$, which corresponds to a minimum central density $\rt_c$ since there is a one-to-one correspondence between these two quantities for hydrodynamically stable stars due to Eq.~\eqref{eq:dmdrho}. However, note that this does not mean all stars above this minimal central density are tachyonically unstable, since the pressure becomes more and more important as densities increase, and $\tilde{T}= -\tilde{\rho}+3\tilde{p}$ can start to decrease for the most massive stars. Because of this, GR solutions above a certain mass may not have a tachyonic instability, meaning there are scalarized stars in a mass range, and we have the unscalarized solutions both below and above this. This is not a necessity, $\tilde{T}$ may never become small enough depending on the EOS and $\beta$. In that case, all stable solutions above a certain mass are scalarized. These results have been directly confirmed in explicit computation of scalarized neutron star structures using the modified TOV equations~\cite{PhysRevLett.70.2220,Ramazanoglu:2016kul,Doneva:2022ewd}. The onset of scalarization is similar in the linear DEF model, which follows the same dependence of scalarization on neutron star mass for $\beta<0$.

The situation is different for $\beta>0$, which is only relevant for the linear DEF model. Recall that scalarization requires $\tilde{T}>0$ now, which can only occur for massive enough stars, and $\tilde{T}$ typically increases monotonically once it changes sign. Hence, there is not an upper limit to $M$ or $\rt_c$ where scalarization stops, and once we determine the least massive star susceptible to the tachyonic instability, we know that all stars with higher mass are also susceptible. We should, however, remember that there is no guarantee for this instability to end in a scalarized neutron star solution.

Lastly, in the sGB model where we only consider $\betagb>0$, $\mathscr{G}$ has a role somewhat analogous to $\tilde{T}$ of the quadratic DEF model with $\beta>0$. That is, $\mathscr{G}$ behaves monotonically with increasing stellar mass ~\cite{Doneva:2017duq}. This means that this model also behaves similarly to the $\beta>0$ case of the linear DEF model in that once the tachyonic instability appears at a certain $M$ (equivalently a certain $\rt_c$) value, all stars above that mass are also unstable to scalarization.

We want to reiterate that the ultimate test for instability is nonlinear time evolution, which would indeed show that the solution evolves away from the initial configuration. Such studies have been limited to very few cases, but linearized mode analysis has confirmed the above results so far~\cite{Doneva:2017duq,Mendes:2018qwo,Doneva:2022ewd}, hence we will assume them to hold in general.

\subsection{Trends on the model parameter spaces}
\label{sec:trends}
%
\begin{figure}
    \includegraphics[width=\columnwidth]{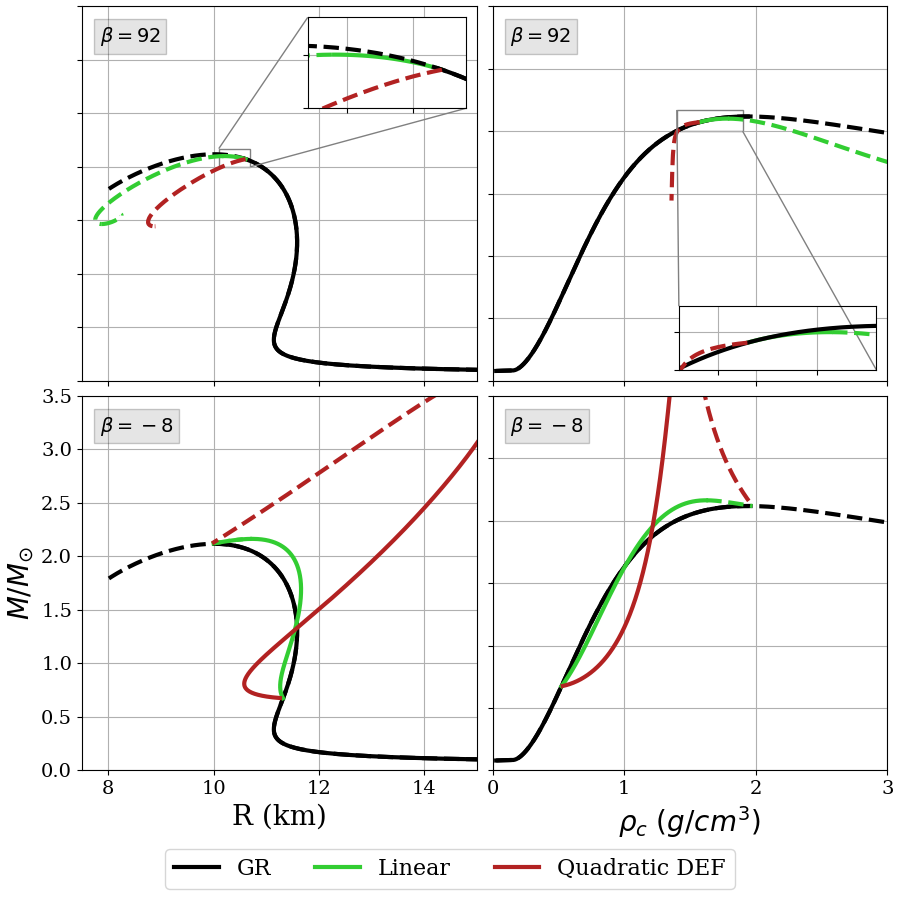}
    \caption{Mass-radius and mass-central density diagrams of the linear (green)and quadratic (red) DEF models for $\beta = -8$ (lower) and $\beta = 92$ (top). HB EOS is used for both cases. Solid lines refer to the hydrodynamically stable solutions and dashed lines to the unstable ones. Deviations from GR for the linear case are much smaller than that of the quadratic one, which is especially pronounced for $\beta<0$.}
    \label{fig:all_DEF}
\end{figure}
\begin{figure}[h]
    \includegraphics[width=\columnwidth]{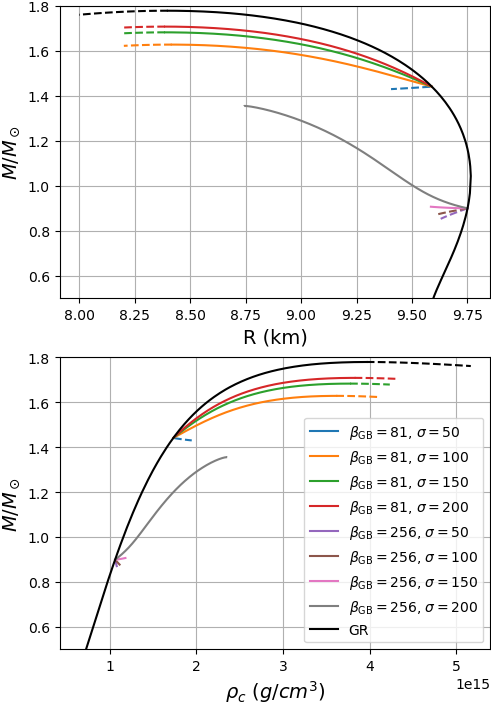}
    \caption{The mass-radius and mass-central density diagrams for different parameter values of sGB theory and for 2B EOS. Hydrodynamically unstable solutions are shown with dashed lines.}
    \label{fig:GB_mass_radius}
\end{figure}

The mass-radius curves we obtained following our recipe can be found in Fig.~\ref{fig:all_DEF} for the linear DEF model and in Fig.~\ref{fig:GB_mass_radius} for the sGB model. We see that there are cases where there is more than one star solution with the same mass $M$, even when GR is assumed. Then, either one of the solutions will be the stable one, and this is the solution we will later use for the Bayesian analysis, or there will be no stable solution at all.\footnote{It is possible that in some cases there are metastable solutions, solutions that are robust against small perturbations, but go to an even more stable configuration with lower total energy if the disturbance is strong~\cite{Tuna:2022qqr}. We will not consider such configurations, but their investigation is an interesting project on its own.}

Let us start with the linear DEF model in Fig.~\ref{fig:all_DEF}, where we plot the curves for both the quadratic and linear DEF models for comparison. Recall that sufficiently negative values of $\meff^2$ are needed for scalarization. Hence, there is an interval of $\beta$ around zero (that also mildly depends on EOS), $\beta_{\rm max} >\beta > \beta_{\rm min}$, where there is no scalarization. Since scalarization is known to change in a continuous manner, we also expect deviations of the scalarized mass-radius curves to be closer to those of GR for low $|\beta|$, and grow as $|\beta| \to \infty$. which is observed in all existing work~\cite{Doneva:2022ewd}. These points are confirmed with numerical computations. 

A radical difference between the quadratic and linear DEF models in Fig.~\ref{fig:all_DEF} is the amount of deviation from GR. For the same $\beta$ value, the linear model always provides smaller deviations when both models have stable solutions. Even without any formal statistical analysis, this is a strong indication that the constraints that can be obtained from mass-radius data on the linear DEF model are weaker than those on the quadratic one, which will be the case in our results.

In light of our instability discussion in Sec.~\ref{sec:stability}, note that for $\beta>0$ any GR solution whose mass is above the maximum of the scalarized branch is also tachyonically unstable, e.g. $\beta=92$ in Fig.~\ref{fig:all_DEF}. Hence, the maximum astrophysically observable neutron star mass in the linear DEF model with $\beta>0$ is always less than the case of GR.

The general trends of the mass-radius curves of the sGB model as we move on the $(\betagb,\sigma)$ parameter space can be seen in Fig.~\ref{fig:GB_mass_radius}. Recall that $\betagb$ alone determines if there is a tachyonic instability or not, hence the onset of scalarization. On the other hand, $\sigma$ is important to determine the structure, e.g. the stellar mass and radius, of the final scalarized star since it controls the coupling  for higher values of $\phi$. This can be clearly seen in the figure. For a given $\betagb$, the mass-radius curves corresponding to different $\sigma$ are different, but they all branch off from the GR curve at the same point, the onset of scalarization.

We know that there is no tachyonic instability, hence no scalarization, below a nonzero value of $\betagb$, see Eq.\eqref{eq:sgb_eom}. Similarly, the nonlinear effects are extremely strong, and suppress the tachyonic instability even for small values of $\phi$ when $\sigma \to \infty$, hence low levels of scalarization and neutron star structures similar to those of GR are also expected in this part of the parameter space. That is, we expect star solutions to approach those of GR in the $\betagb \to 0$ and $\sigma \to \infty$ limits. Both trends are apparent in Fig.~\ref{fig:GB_mass_radius}.

Instability considerations render some parts of the mass-radius curves astrophysically  irrelevant for both models. One prominent example is the $\sigma \to 0$ limit in the sGB model. For any given value of $\betagb$ in Fig.~\ref{fig:GB_mass_radius}, the scalarized branch satisfies $dM/d\tilde{\rho}_c <0$ \emph{everywhere} if $\sigma$ is below a certain critical value which we will denote with $\sigmacrit$. This means all such solutions are hydrodynamically unstable due to Eq.~\eqref{eq:dmdrho}:
\begin{align} \label{eq:simga_crit}
  \sigma < \sigmacrit\ \Rightarrow\ \textrm{ all scalarized stars are unstable} .
\end{align}
So, even though we can find equilibrium scalarized star solutions for $\sigma < \sigmacrit$, these are not astrophysically relevant, and will not be part of our statistical analysis. 

Eq.~\eqref{eq:simga_crit} has drastic consequences for the sGB model in light of our stability discussion. Any star that has a mass above where the scalarized branch leaves the GR branch is unstable. For example, when $\betagb=256$, $\sigma<100$ (see Fig.~\ref{fig:GB_mass_radius}), we would only have stable neutron stars with $M \lesssim 0.9 M_\odot$. Moreover, when there are stable scalarized stars, any GR solution whose mass is above the most massive stable scalarized solution is also unstable, as in the linear DEF model. One example is the GR solutions (black line) with $M>1.37 M_\odot$ in the case of $\betagb=256$, $\sigma=200$ in Fig.~\ref{fig:GB_mass_radius}. These points will be essential in our Bayesian likelihood computation.

\subsection{Bayesian analysis}
\label{sec:bayesian_analysis}
Our statistical approach also follows \textcite{Tuna:2022qqr}, who in turn followed \textcite{Ozel:2015fia}. To summarize, we use Bayesian inference to obtain posterior likelihoods for a set of parameters $\{\pi\}$, using the mass-radius measurements of $N=16$ stars. {14 of these (12 from \textcite{Ozel:2015fia}, 2 from \textcite{Bogdanov:2016nle}) are the same ones used in the analysis of \textcite{Tuna:2022qqr}. The two new stars are PSR J0030+0451~\cite{j0030_miller,j0030_riley} and PSR J0740+6620~\cite{j0740_miller,j0740_riley}, each of which was analyzed by two separate research groups. All stars have equal weight in our prior, and we assign a half weight for the mass-radius data obtained by each team for the two new stars}.\footnote{{There is also other work which provides upper or lower bounds on the mass and radii of neutron stars~\cite{Bauswein:2017vtn_constrain1,Dietrich:2020efo_constrain2,Annala:2017llu_constrain3,Capano:2019eae_constrain4}, but we could} {not directly incorporate these to our analysis since they did not provide explicit probability distributions for the stellar mass and radius.}} 

The set of our model parameters can be defined as
\begin{align}
    \{\pi\} \equiv \{\beta_n, \lambda\}
\end{align}
where $\lambda$ stands for the EOS and $\beta_n$'s are the dimensionless parameters of each model we investigate, e.g., $\{\pi_{GB}\} = \{\beta_{GB}, \sigma, \lambda\}$ for the sGB model. 

Each mass-radius measurement in the dataset we used is a likelihood distribution $P_i(M,R)$ with $i=1,\dots, N$. Therefore, to obtain the posterior likelihood of a certain set of parameters given the data, $P(\{\pi\}|\text{data})$, we use Bayes' theorem
\begin{align}
    P(\{\pi\}|\text{data}) = \mathcal{N}P(\text{data}|\{\pi\})\text{P}_{\text{Pri}}(\{\pi\})
\end{align}
where  $ \mathcal{N}$ is the normalization constant and $\text{P}_{\text{Pri}}(\{\pi\})$ is the prior likelihood of the parameters. We can also write the likelihood of the dataset as a product of the likelihood of each individual data point
\begin{align}
    P(\text{data}|\{\pi\}) = \prod_{i=1}^N  P_i(M,R|\{\pi\})\ ,
\end{align}
where we used the fact that each data point corresponds to the probability distribution of a neutron star mass-radius measurement, $P_i(M,R)$, as we discussed earlier. 

Since each neutron star mass-radius curve can be thought of as a one-to-one function of mass, i.e., for a given set of parameters $\{\pi\}$ one can write $R = R(M;\{\pi\})$, it is possible to turn the expression above into an integral by marginalizing over $M$ as
\begin{align}\label{eq:M_integaral}
    P_i(M,R|\{\pi\})=\int^{M_{max}}_{M_{min}}P_i(M,R(M;&\{\pi\}) \nonumber\\ \times &\text{P}_\text{Pri}(M)dM.
\end{align}
That is, the likelihood of an individual neutron star mass-radius measurement for a given set of parameters $\{\pi\}$ can be calculated by integrating over the mass-radius curve determined by $R(M;\{\pi\})$. We take the prior likelihood of the mass of the neutron star $\text{P}_\text{Pri}(M)$ to be flat as has been the case in the literature~\cite{Tuna:2022qqr,Ozel:2015fia}.

\begin{figure}
    \includegraphics[width=\columnwidth]{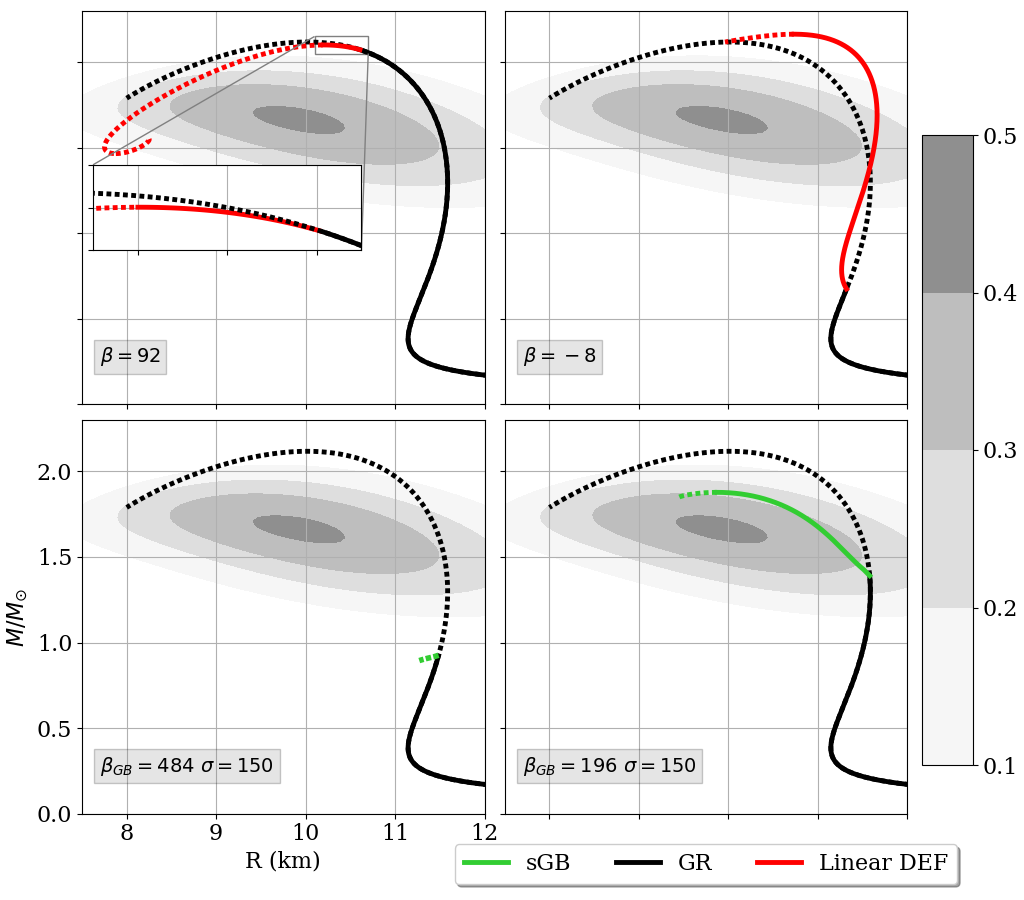}
    \caption{Mass-radius curves for HB EOS for various scalarization models, together with the contours of the mass-radius probability distribution for the star KS1731-260. Stable stars that are compared to the mass-radius measurements for our Bayesian analysis are solid, and the unstable ones due to hydrodynamics or scalarization are dotted. Upper left: Linear DEF model with $\beta= 92$ where the GR solutions above the maximum mass of the scalarized branch are unstable to scalarization, hence the maximum mass is lower than the GR case. Upper right: Linear DEF model with $\beta=-8$ where the maximum mass is on the scalarized branch, and is higher than the GR case. Lower right: sGB model with $\betagb=196$, $\sigma=150$ where the maximum mass is lower than the GR case and is on the scalarized branch, like the upper left. Lower left:  sGB model with $\betagb=484$, $\sigma=150$ where the scalarized branch is completely hydrodynamically unstable and GR stars above the point of branch-off are unstable to scalarization, hence, we only have a smaller portion of the GR mass-radius curve.
    }
   \label{fig:mr_curves_and_probability}
\end{figure}

Recall that there is at most one stable star for a given mass value $M$, which means all astrophysically relevant solutions will form a single parameter family, the parameter being $M$. Following our stellar stability discussion in the previous section, the theoretical mass-radius curves can have various different structures, some of which we present in Fig.~\ref{fig:mr_curves_and_probability} juxtaposed with the mass-radius distribution for one of the stars in our dataset. 

Five different piecewise-polytropic EOS are used to demonstrate the behavior of the nuclear matter: 2H, H, HB, B, 2B introduced in \textcite{2009PhRvD..79l4033R}. They cover a wide range of physical possibilities ranging from the very soft (2B) to the stiffest (2H), which sufficiently serves our purpose of observing the leading effects of EOS on constraining scalarization models. We will see that 2H EOS will contribute insignificantly to the posterior likelihood, the underlying reason being the radically large masses and radii it predicts as can be seen in Fig.~\ref{fig:gr_mr}. These mass-radius curves have little to no overlap with the mass-radius data of some of the stars (M28, X5, and X7 of \textcite{Ozel:2015fia}) even after scalarization effects are included, hence the low posterior likelihood. Nevertheless, we kept all five EOS for the sake of completeness.

As for the model parameters, we will have two separate prior choices. First, we use flat prior distributions in all cases: $\beta \in [-202,\ 212]$  for the linear DEF model; $\betagb \in [36,\ 900]$, $\sigma \in [50,\ 700]$ for the sGB model. Together with a flat prior on the EOS parameter, we call this the \emph{simple flat prior}. These values cover all the parameter ranges of interest in the literature so far. 

{The above limits on the priors can seem excessively large at first sight. Even though we are not familiar with a concrete study on the specific models we are going to investigate, the binary observations that ruled out the original DEF model might provide some bounds on our prior parameter space as well, which might suggest considering a smaller region of the parameter space for the prior. However, large parameter values such as $|\beta|=200$ and $\betagb =900$ are relevant when we consider massive scalars. The theories we investigate here have massless scalars, but scalarization is known to continuously change with the scalar mass $m_\phi$, which  means the structure of a scalarized neutron star where $m_\phi=0$ is practically identical to a star where $m_\phi$ is very low~\cite{Ramazanoglu:2016kul,Tuna:2022qqr}. Thus, our results are informative for $m_\phi \neq 0$ as well. The crucial point for the limits of our parameters is that, a scalar mass term renders the binary observations irrelevant similarly to the case of the quadratic DEF model~\cite{Ramazanoglu:2016kul,Tuna:2022qqr}, hence, any theory parameter value becomes viable again. Overall,  in the prior, we chose to consider the large parameter space regions in the above paragraph, so that we can probe the parameter values that are not yet ruled out for $m_\phi \neq 0$.}

The simple flat prior has some subtle points for the sGB model. This model is known to have purely theoretical well-posedness problems for high values of $\betagb$, where the theory may not be mathematically meaningful to start with~\cite{East:2021bqk,Corman:2022xqg,R:2022hlf}.\footnote{{\textcite{Minamitsuji:2023uyb} is posted as a preprint while we are revising this manuscript, which might also lead to some constraints based on the stability of the solutions in the sGB theory.}} Hence, using this finite interval for $\betagb$ is well motivated. As for $\sigma$, recall that the neutron star structures become identical to those of GR when $\sigma \to \infty$. This means, there is an infinitely large part of the $(\betagb, \sigma)$ parameter space where the posterior is identical to that of GR, and this part can artificially dominate the Bayesian analysis if arbitrarily large values of $\sigma$ are considered. One would like to avoid these parts of the parameter space where the theory is identical to GR even though it uses a complicated machinery with extra parameters. One potential choice is weighting the prior so that this infinite-but-useless region is punished, such as $e^{-\kappa \sigma}$ with $\kappa>0$. We made the simpler choice of a sharp cutoff at the quite large value of $\sigma=700$ as a conservative approach.\footnote{\textcite{Tuna:2022qqr} encountered a similar issue for one of their parameters, the scalar mass $\mphi$, and showed that using different priors to avoid the infinitely large region with nonzero posterior provides qualitatively similar constraints.} Also note that the natural expectation is that the dimensionless parameter $\sigma$ is not much larger than unity, which further supports a finite range for $\sigma$.

Our second prior on the model parameters is simply restricting the simple flat prior to the cases where the maximum theoretically possible neutron star mass is at least two solar masses, $2M_\odot$, which is known through separate observations~\cite{Demorest:2010bx,Romani:2022jhd}. We call this the \emph{restricted prior}. Factors such as high spin increase the theoretical maximum stellar mass, and some mass measurement techniques have relatively high error bars, hence we picked this relatively conservative value as our cutoff. 2B and B EOS under GR are already ruled out in this prior, see Fig.~\ref{fig:gr_mr}. This does not mean that considering these EOS is meaningless, since the maximum theoretically allowed mass for a neutron star under scalarization can increase compared to GR, linear DEF model with $\beta<0$ in Fig.~\ref{fig:mr_curves_and_probability} being a well-known example. Nevertheless, we will see that scalarization will not change the GR picture, and these EOS will be practically eliminated in the posterior likelihood for the restricted prior.

In summary, the simple flat prior represents an analysis that purely relies on the simultaneous mass-radius measurements from the x-ray data, whereas the restricted prior incorporates additional observations. We will see that both priors lead to qualitatively similar conclusions.

\section{Results}
\label{sec:results}
\subsection{Extensions of the DEF model}
\label{sec:def_model}
We first perform the Bayesian analysis of the previous section on the linear DEF model to obtain the posterior probability distribution $P(\beta,\lambda|\mbox{data})$. The \emph{unnormalized} conditional probability for each EOS $\lambda$
\begin{align}\label{eq:linearDEF_prob_cond}
    \begin{aligned}
    P^\lambda_{\mbox{\scriptsize{cond}}}(\beta) &\equiv P(\beta,\lambda|\mbox{data}) \, , 
    \end{aligned}
\end{align}
and the marginal probability 
\begin{align}\label{eq:linearDEF_prob_unmarg}
    \begin{aligned}
    P_{\mbox{\scriptsize{marg}}}(\beta) &\equiv \sum_\lambda P(\beta,\lambda|\mbox{data}) = \sum_\lambda P^\lambda_{\mbox{\scriptsize{cond}}}(\beta)
    \end{aligned}
\end{align}
are useful in interpreting the posterior. The former tells us the behavior when we assume each of the EOS, hence comparing these provide information about how much our results depend on nuclear physics. It also shows the relative probability of each EOS in the posterior since we do not individually normalize $P^\lambda_{\mbox{\scriptsize{cond}}}(\beta)$. $P_{\mbox{\scriptsize{marg}}}(\beta)$ marginalizes the EOS parameter which covers a quite large range, hence it is a measure of the likelihood of $\beta$ where the effects from EOS is averaged out.

\begin{figure}[t]
\includegraphics[width=\columnwidth]{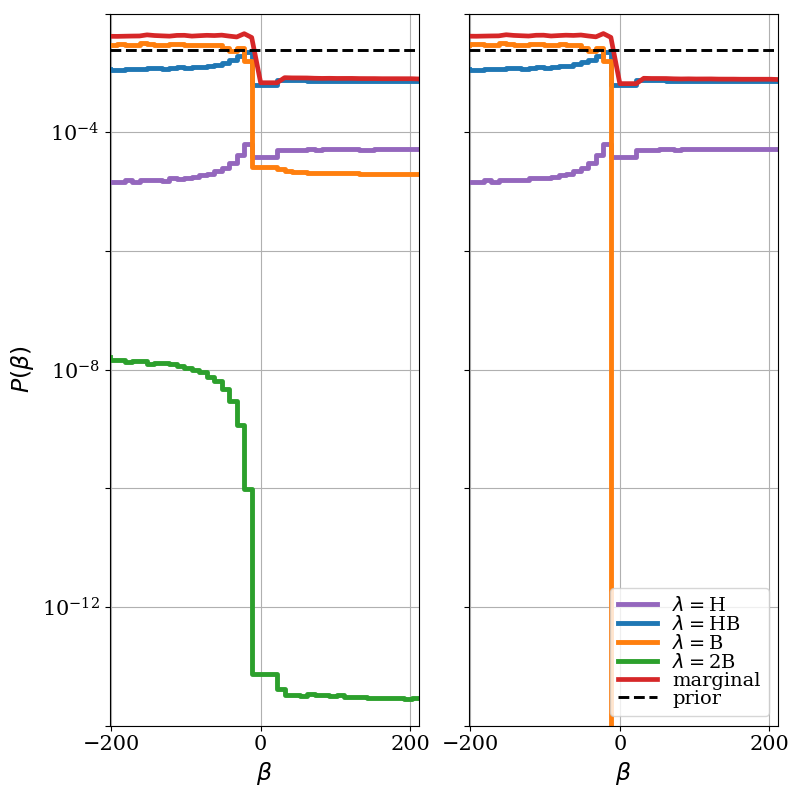}
\caption{Conditional posterior probabilities for the linear DEF model using the simple flat prior (left) and the restricted prior (right). Individual lines are the unnormalized conditional probabilities for each EOS in Eq.~\eqref{eq:linearDEF_prob_cond} and the marginal probability in Eq.~\eqref{eq:linearDEF_prob_unmarg}. {2H EOS is not shown as it has a negligible contribution to the posterior for $\beta<0$ and vanishes for $\beta>0$.} }
\label{fig:linearDEFposterior}
\end{figure}
The resulting posterior distributions can be seen in Fig.~\ref{fig:linearDEFposterior} for both of our prior choices. The restricted prior simply rules out 2B EOS completely, and B EOS for $\beta>0$. We will discuss some of the differences in the results provided by the two priors, but they both lead to the same major conclusion.

The posterior likelihoods are not completely featureless in Fig.~\ref{fig:linearDEFposterior}, meaning the mass-radius data provides some information about what parts of the parameter space are more likely than the others. For example, looking at the flat prior computation, there is a strong preference for $\beta<0$ if 2B EOS is assumed, {which is also the case for B and HB EOS, albeit less prominently.} However, we see that this is not the case for all EOS, e.g. the opposite is true for {H} EOS, and the contrast between $\beta \lessgtr 0$ is much lower. {There isn't a significant change when we use the restricted prior, apart from the fact that the posterior for B EOS for $\beta>0$ and 2B EOS for all $\beta$ vanishes.} 

Overall, the posterior averaged over the EOS, the marginal distribution, deviates from the prior {only by a limited amount both for the flat and restricted priors. Hence, the mass-radius data is not particularly useful for constraining this model.}

The most important aspect of the posterior distributions for both prior choices is that the probability densities do not decay in any discernible manner as $\beta \to \pm \infty$. Hence, the posterior might still be large outside of the $\beta$ range we studied, in principle even for arbitrarily large $\beta$. This means our inferences from these curves are strongly dependent on the range of $\beta$ values we use, and we cannot reliably conclude $\beta$ to have a lower or upper bound based on mass-radius measurements alone.

Recall that we already predicted that the linear DEF model would be less stringently constrained compared to the quadratic DEF model in Sec.~\ref{sec:trends}, due to the fact that its deviations from GR are smaller. Here, this is confirmed in the most extreme way: There is no reliable bound on $\beta$ at all. This is the first major result of this study.

This is also a reminder that the common lore of spontaneous scalarization leading to large deviations from GR can sometimes be misleading. The switch from the quadratic DEF model to the linear one enabled the formation of stable scalarized neutron stars for $\beta>0$, and this is perhaps related to the fact that the deviations from GR are somewhat suppressed for the linear DEF model.

\subsection{Scalar-Gauss-Bonnet theories}
\label{sec:def_model}
The investigation of the scalar-Gauss-Bonnet theory is similar to that of the DEF model aside from having two theory parameters $(\beta_{\rm GB}, \sigma)$ in Eq.~\eqref{eq:sgb_action} and~\eqref{eq:f_parametrized}. As before, we define the conditional and marginal posterior probabilities related to the EOS as
\begin{align}\label{eq:GB_prob_unmarg}
    \begin{aligned}
    P^\lambda_{\mbox{\scriptsize{cond}}}(\beta_{\rm GB},\sigma) &\equiv P(\beta_{\rm GB},\sigma,\lambda|\mbox{data}) \\
    P_{\mbox{\scriptsize{marg}}}(\beta_{\rm GB},\sigma) &\equiv \sum_\lambda P(\beta_{\rm GB},\sigma,\lambda|\mbox{data}) \ ,
    \end{aligned}
\end{align}
respectively. We also further marginalize over $\betagb$ or $\sigma$ to obtain the one-parameter probability distributions 
\begin{align}\label{eq:GB_prob_marg_lambda}
    \begin{aligned}
    P^\lambda{\mbox{\scriptsize{cond}}}(\betagb) &\equiv \int_0^\infty d\sigma \, P^\lambda_{\mbox{\scriptsize{cond}}}(\beta_{\rm GB},\sigma) \, \\
    P_{\mbox{\scriptsize{marg}}}(\betagb) &\equiv \int_0^\infty d\sigma \, P_{\mbox{\scriptsize{marg}}}(\beta_{\rm GB},\sigma)) \\
    P^\lambda{\mbox{\scriptsize{cond}}}(\sigma) &\equiv \int_0^\infty d\betagb \, P^\lambda_{\mbox{\scriptsize{cond}}}(\beta_{\rm GB},\sigma) \\
    P_{\mbox{\scriptsize{marg}}}(\sigma) &\equiv \int_0^\infty d\betagb \, P_{\mbox{\scriptsize{marg}}}(\beta_{\rm GB},\sigma)\ .
    \end{aligned}
\end{align}

\begin{figure}[t]
    \includegraphics[width=\columnwidth]{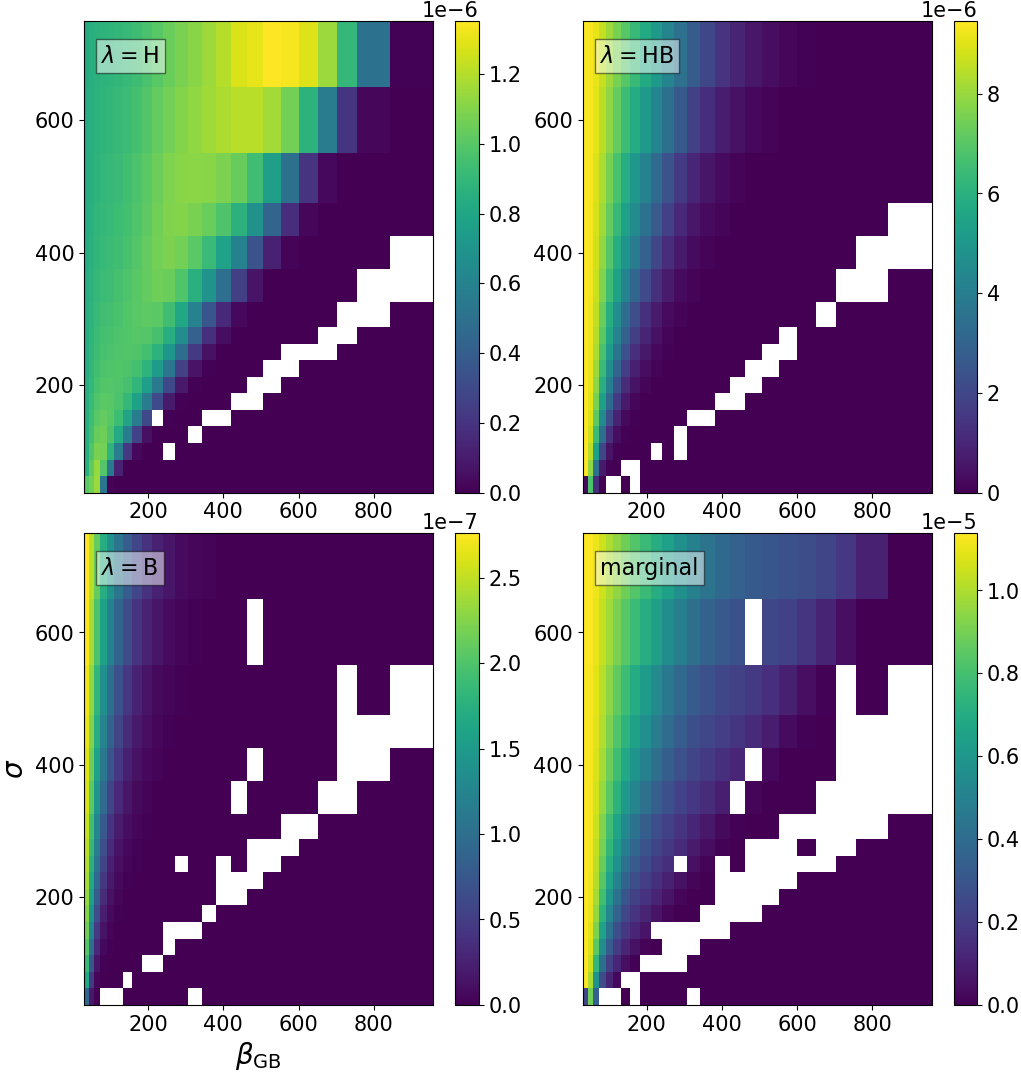}
    \caption{Conditional and marginal posterior probability densities for the sGB model (Eq.~\eqref{eq:GB_prob_unmarg}). We only show H, HB, and B EOS since the other two contribute negligibly to the posterior, {or do not contribute at all.} Each subfigure has a separate color scale in order to highlight its details. The blank cells are due to the computational failures we encountered around the region where the stable scalarized solutions switch to unstable ones near $\sigmacrit$. 
    }
    \label{fig:GB_3EOS_heatMap}
\end{figure}
Let us first start with Fig.~\ref{fig:GB_3EOS_heatMap} where you can find the posterior distributions on the $(\betagb, \sigma)$ parameter space for the simple flat prior. Note that all the trends we mentioned in Sec.~\ref{sec:trends} are followed, for example, $\betagb \to 0$ region has the same likelihood, which is the likelihood for the GR mass-radius curve. For, $\sigma<\sigmacrit$, the likelihood is strictly less than that of GR since the mass-radius curve is simply a subsection of the GR curve. The posterior probabilities in this region is also ignorably small or simply zero, especially for large $\betagb$. This is because the mass-radius curves have an exceedingly small overlap with the data.


%
\begin{figure}
\includegraphics[width=\columnwidth]{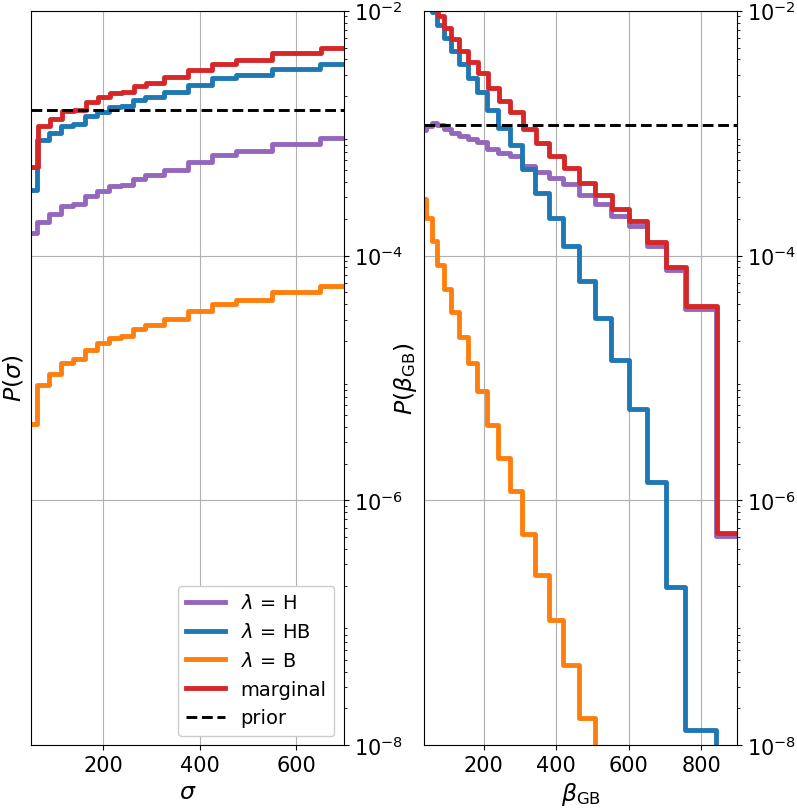}
\caption{Conditional and marginal posterior probability densities as a function of only one of $\betagb$ or $\sigma$ for the sGB model, Eq.~\eqref{eq:GB_prob_marg_lambda}, using the simple flat prior. {2B EOS has the same qualitative behavior as the EOS we show, though the probability values are too low to appear in the plot.} 2H EOS has a vanishing likelihood as explained in the text. The exponential, or faster, decays at high $\betagb$ (right) and low $\sigma$ (left) lead to the bounds in Eq.~\eqref{eq:bound_sgb_prior1} and Table.~\ref{GB_EOS_bounds}.}
\label{fig:GB_marg}
\end{figure}
Some of the features of our Bayesian analysis can be more easily observed in Fig.~\ref{fig:GB_marg}, where we individually marginalize over each of $\betagb$ and $\sigma$ as in Eq.~\eqref{eq:GB_prob_marg_lambda}.\footnote{We filled the blank parts of the parameter space in Fig.~\ref{fig:GB_3EOS_heatMap} using interpolation from the neighboring points before performing the integrals needed for the marginal probabilities in Eq.~\eqref{eq:GB_prob_marg_lambda}.} When EOS is averaged out, $P{\mbox{\scriptsize{marg}}}(\betagb)$ and $P{\mbox{\scriptsize{marg}}}(\sigma)$, we see that the probability distribution decays exponentially or faster at high $\betagb$ values and low $\sigma$ values. Taken at face value, this suggests that we can find bounds on these parameters. For the simple flat prior
\begin{align}\label{eq:bound_sgb_prior1}
    \begin{aligned}
        \betagb &< 491  \\
        \sigma &> 162
    \end{aligned}
\end{align}
at $95\%$ confidence based on the marginalized distributions. {We also looked at how sensitive these bounds are to the EOS. Table.~\ref{GB_EOS_bounds} lists the bounds if we separately assume each EOS. }
\begin{table}[h!]
\begin{tabular}{|c || c | c|} 
 \hline
EOS & \multicolumn{2}{c|}{95\% bounds} \\ [0.5ex] 
 \hline\hline
H\ &$ \sigma > 175$ &  $\ \betagb < 635$  \\[0.5ex]
HB \ &$ \sigma > 158$  & $\ \betagb < 318$\\[0.5ex]
B \ &$ \sigma > 176$  & $\ \betagb < 187$ \\[0.5ex]
2B\ &$ \sigma > 308$ & $\ \betagb < 67$  \\[0.5ex]
 \hline
\end{tabular}
\caption{{95\% confidence bounds for the sGB model with the flat prior when we assume each EOS separately. There is no clear trend for $\sigma$, but $\betagb$ is more easily constrained for softer EOS.}}
\label{GB_EOS_bounds}
\end{table}
%


Our ability to obtain the bounds in Eq.\eqref{eq:bound_sgb_prior1} and {Table.~\ref{GB_EOS_bounds}} is in line with the observations in Secs.~\ref{sec:trends} and~\ref{sec:bayesian_analysis}. Recall that we deviate further from the GR curves as $\betagb \to \infty$, and the mass-radius curves have exceedingly small maximum stellar masses, hence the overlap with data becomes worse, see Fig.~\ref{fig:GB_mass_radius}. This means the posterior probability was expected to go down in this part of the parameter space. At the $\sigma \to 0$ limit, all scalarized stars tend to be hydrodynamically unstable, and the GR solutions are tachyonically unstable to scalar growth. Thus, the mass-radius curve is only a smaller portion of what we have in GR, see Fig.~\ref{fig:mr_curves_and_probability}. As this portion gets smaller, we also have less overlap with data, and small $\sigma$ values become extremely unlikely or they are outright ruled out. 

Despite the agreement with expectations, there are also reasons to be cautious about the $\betagb \lesssim 500$ bound of Eq.\eqref{eq:bound_sgb_prior1}. First, observe that the dependence of the posterior distribution on $\betagb$ differs between EOS. {In Fig.~\ref{fig:GB_marg}, there is exponential decay in the $\betagb \to \infty$ limit in all cases, however the rate of decay varies between EOS, constraints on softer EOS being stronger (see Table.~\ref{GB_EOS_bounds}).} Nevertheless, the fact that a clear upper $\betagb$ bound exists for any EOS demonstrates the power of the mass-radius data in constraining this model. {There is also some variance for the bounds on $\sigma$ depending on the EOS, however, there is no clear trend. Moreover, the three EOS that contribute the most to the posterior (H, HB, and B) provide very similar bounds.}



There is also some ambiguity in the choice of the dimensionless parameters $\betagb$ and $\sigma$. Our results provide bounds on generic functions $f(\betagb, \sigma)$ as well, but using an alternative parameter could affect the results through the prior choice. For example, if $\sqrt{\betagb}$ is used as a primary parameter as in \textcite{Doneva:2017duq}, a flat prior would be equivalent to a prior on $\betagb$ that is weighted as $\betagb^{-1/2}$. This leads to an even more stringent constraint, i.e. a lower upper bound for $\betagb$ than Eq.~\eqref{eq:bound_sgb_prior1}, but other choices might lead to the opposite effect of having weaker constraints.

\begin{figure}[h]
\includegraphics[width=\columnwidth]{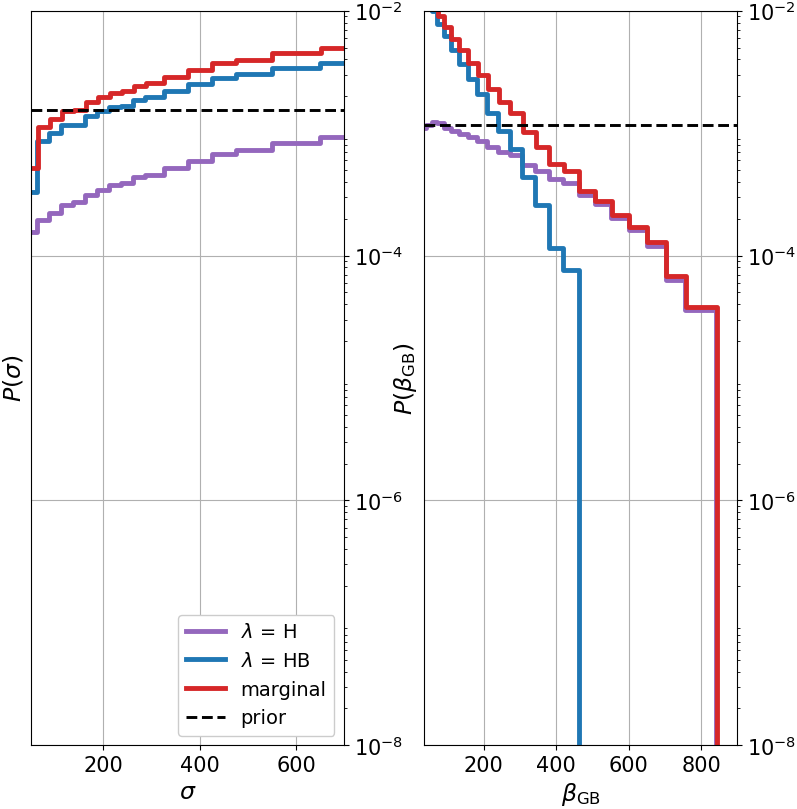}
\caption{Same as Fig.~\ref{fig:GB_marg}, but for the restricted prior. For 2B and B EOS, the maximum theoretically allowed mass in the sGB model is less than $2M_\odot$, hence they have vanishing posterior likelihood. Recalling that 2H EOS also has vanishing posterior likelihood, this leaves only H and HB EOS as viable choices.}
\label{fig:GB_marg_restricted}
\end{figure}
The results for our other prior choice, the restricted prior, are given in Fig.~\ref{fig:GB_marg_restricted}, which lead to the bounds
\begin{align}\label{eq:bound_sgb_prior2}
    \begin{aligned}
        \betagb &< 427  \\
        \sigma &> 133
    \end{aligned}
\end{align}
at $95\%$ confidence based on the marginalized distributions. These are slightly more stringent than the case of the flat prior in Eq.~\eqref{eq:bound_sgb_prior1}. However, the behavior is qualitatively similar aside from the fact that the softer 2B and B EOS are now completely ruled out.

Overall, the bound on $\betagb$ of the sGB model is less stringent than the one on $\beta$ of the quadratic DEF model obtained in \textcite{Tuna:2022qqr} (which is $\beta \gtrsim -20$ where only $\beta<0$ was investigated). Very high values of $\betagb$ are disfavored, but it is hard to say whether this should be in the lower or higher hundreds. {Hence, considering the variation with the unknown EOS, it is more appropriate to report a softer bound of $\betagb \lesssim 500$, as we did above.} This is weaker than the bounds that might be obtained by hyperbolicity considerations, whose interpretation is still under discussion~\cite{East:2021bqk,Corman:2022xqg,R:2022hlf}. In any case, our results demonstrate the utility of the mass-radius data as a tool to test gravity theories.

As for $\sigma$, there is already a natural mechanism to restrict it to low values as we discussed in relation to Fig.~\ref{fig:mr_curves_and_probability}. Even though order of unity changes are possible due to the details of our methodology, we consider a soft bound of $\sigma \gtrsim 100$ to be clearly the case.

\section{Effect of the dataset size}
\label{sec:dataset_size}
%
\begin{figure}[t]
    \includegraphics[width=\columnwidth]{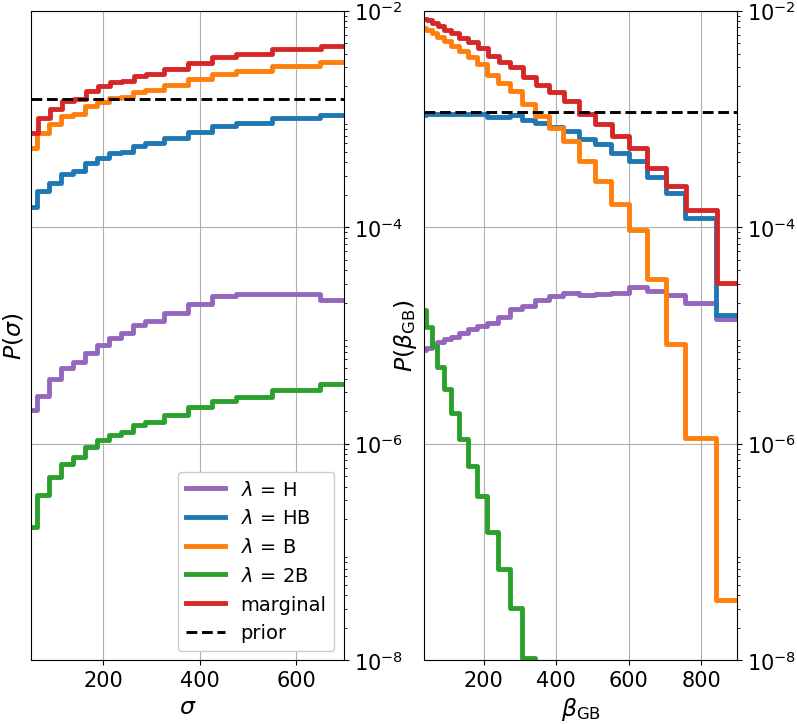}
    \caption{Same as Fig.~\ref{fig:GB_marg} but for the smaller dataset that excludes the stars PSR J0030+0451 and PSR J0740+6620. The most notable change is for H EOS, where the decay for $P(\betagb)$ barely starts within the parameter range we consider. This is unlike Fig.~\ref{fig:GB_marg} where all EOS have clear exponential decay at high $\betagb$.  }
\label{fig:GB_marg_old}
\end{figure}
\begin{table}[h!]
    \begin{tabular}{|c || c | c|} 
    \hline
    EOS & \multicolumn{2}{c|}{95\% bounds} \\ [0.5ex] 
    \hline\hline
    H \ &$ \sigma > 217$ & $\ \betagb < 859$  \\[0.5ex]
    HB \ &$ \sigma > 180$ & $\ \betagb < 697$  \\[0.5ex]
    B \ &$ \sigma > 162$ & $\ \betagb < 465$  \\[0.5ex]
    2B \ &$ \sigma > 191$ & $\ \betagb < 166$  \\[0.5ex]
    \hline
    \end{tabular}
    \caption{Same as Table.~\ref{GB_EOS_bounds}, but for the smaller dataset that excludes the stars PSR J0030+0451 and PSR J0740+6620. All $\betagb$ bounds are weaker compared to the bigger dataset, as expected, though, there is no clear trend for $\sigma$. The $\betagb$ bound for the H EOS is quite close to the highest $\betagb$ value we consider, hence it is debatable whether there is a bound at all.}
    \label{GB_EOS_bounds_old}
\end{table}
The number of stars whose mass-radius measurements we use also affects our ability to constrain scalarization. In this study, we used data from two more recently observed stars that were not part of the similar study of \textcite{Tuna:2022qqr}: PSR J0030+0451~\cite{j0030_miller,j0030_riley} and PSR J0740+6620~\cite{j0740_miller,j0740_riley}. In order to have an informed opinion about the dependence of our findings on the number of studied stars, we repeated our analysis for a smaller dataset, where the two newer stars are excluded. The results for the sGB model and the flat prior are in 
Fig.~\ref{fig:GB_marg_old} and Table.~\ref{GB_EOS_bounds_old}. When we marginalize over all EOS, we obtain the bounds
\begin{align}\label{eq:bound_sgb_prior1_old}
    \begin{aligned}
        \betagb &< 606  \\
        \sigma &> 168
    \end{aligned}
\end{align}
at $95\%$ confidence. The trends are similar for the restricted prior which we do not present explicitly.

How to interpret these results? First, the $\betagb$ bounds in Eq.~\ref{eq:bound_sgb_prior1_old}, as well as the ones for individual EOS in Table.~\ref{GB_EOS_bounds_old}, are all weaker for the smaller dataset. This is expected, however, the improvement is arguably noteworthy for a seemingly small change in the number of stars from 14 to 16. That is, when we compare Table.~\ref{GB_EOS_bounds_old} with Table.~\ref{GB_EOS_bounds} and Eq.~\ref{eq:bound_sgb_prior1_old} with Eq.~\ref{eq:bound_sgb_prior1}, $\betagb$ is constrained significantly better, even though the change in the amount of data is modest. There isn't such a clear trend for the bounds on $\sigma$, which might be related to the fact that the behavior of the $\sigma$ bounds vary much less prominently with the EOS, e.g. see the bounds in Tables.~\ref{GB_EOS_bounds} or~\ref{GB_EOS_bounds_old}.

Second, and perhaps more significantly, there are also some arguably qualitative changes in our ability to constrain the theory. The behavior of the posterior was the same for all EOS in Fig.~\ref{fig:GB_marg}, there is exponential decay at high $\betagb$. Thus, we could constrain the sGB model for any EOS. We see \emph{mostly} the same trend for the smaller dataset in Fig.~\ref{fig:GB_marg_old}, however, there is the exception of the H EOS where the exponential decay barely starts within our parameter range. Hence, it is arguable whether there is a meaningful upper bound for $\betagb$ for H EOS in Fig.~\ref{fig:GB_marg}. If the smaller dataset was the only one available, we would have to take the bounds in Eq.~\ref{eq:bound_sgb_prior1_old} with a grain of salt.

There was no noteworthy difference between using the smaller or the larger dataset in terms of constraining the linear DEF model. Nevertheless, the improvements we saw for the sGB model are an indication that future mass-radius measurements might have the potential to put statistically significant bounds on this model as well. 

In summary, the simple exercise in this section demonstrates the importance of expanding the observational efforts regarding neutron star mass-radius measurements, and the potential of such endeavors in testing alternative theories.


%

\section{Discussion}
\label{sec:conclusions}
We investigated the general effectiveness of neutron star mass-radius measurements in testing theories of gravity. Other alternative theories have already been constrained in this manner, namely the original spontaneous scalarization model of Damour and Esposito-Farèse, which we called the quadratic DEF model, and its massive scalar extension. We have seen that this model is not an exception, and the sGB model can also be constrained using the same method, albeit more weakly. On the other hand, we could not constrain a modified version of the original DEF model, the linear DEF model. Together, these show that the mass-radius data is a promising tool for testing alternative theories of gravity, but it also has clear limits.

The deviations from GR in the linear DEF model are smaller in comparison to those in the quadratic DEF model, which has an essential role in the ineffectiveness of our methods. We should note that the issue is not trivial. Even though the deviations are smaller, they are not small in absolute terms. The mass-radius curves of the linear DEF model can deviate from those of GR at $\sim 10\%$ level as can be seen in Fig.~\ref{fig:all_DEF}. 

Perhaps the more important difference between the quadratic and the linear DEF models is that even though there is a continuous trend of growing deviations from GR with $\beta \to -\infty$ in the quadratic model, the deviations saturate in the same limit for the linear DEF model. That is, in the linear DEF model, increasing $\beta$ does not change the mass-radius curve beyond a certain point, and the maximum deviations from GR in the $\beta \to \infty$ limit cannot be statistically distinguished from zero due to the uncertainties in the dataset and our methods. 

A rough reasoning for the difference of the two DEF models can perhaps be seen in the behavior of the function
\begin{equation}
    \alpha(\phi) \equiv \frac{d \ln(A)}{d\phi} = 
    \begin{cases}
    \beta{\phi} & \textrm{ quadratic DEF}\\
    \frac{1}{\sqrt{3}} \tanh \left(\sqrt{3}\beta \phi \right)               & \textrm{ linear DEF}\ 
\end{cases} .
\end{equation}
$\alpha(\phi)$ has an important role in the dynamics of the scalar field around any solution, including scalarized ones~\cite{Mendes:2016fby}. It grows without bound in absolute value as $\phi$ grows in the quadratic DEF model, but saturates to the finite value $1/\sqrt{3}$ for the linear case. 

Notwithstanding the details, the above observations suggest that our null result for the linear DEF model may not be essential, and more precise data or methodology can, in principle, lead to constraints on the linear DEF model in the future.

Our ability to constrain the sGB model is important for demonstrating that the quadratic DEF model is not an exception for the use of mass-radius data in testing gravity. The underlying reason for the bounds we obtained is similar to the case of the quadratic DEF model, there are large enough deviations from GR in both cases. However, there is also the difference that whereas the mass-radius curves of the quadratic DEF model deviate mostly in the medium stellar masses and above (Fig.~\ref{fig:all_DEF}), the case is opposite for the sGB model (Fig~\ref{fig:GB_mass_radius}). High $\betagb$ and low $\sigma$ values can be ruled out mainly because the corresponding mass-radius curves do not reach even moderately high stellar masses.

The relationship between the bounds we obtained and those that come from separate tests is also an important issue. The most stringent current results regarding the theories that feature spontaneous scalarization come from the binary star tests for the quadratic DEF model~\cite{2013Sci...340..448A,Freire:2012mg,Zhao:2022vig}. These bounds are based on the scalar radiation from the orbiting stars, which is expected to have a large contribution to the orbital decay~\cite{Doneva:2022ewd}. Hence, the lack of such orbital decay can be used to constrain scalarization, and is claimed to completely rule out the quadratic DEF model~\cite{Zhao:2022vig}. However, even a tiny scalar mass renders binary tests irrelevant by modifying the radiation characteristics~\cite{Ramazanoglu:2016kul, Doneva:2022ewd}. Indeed, the bounds from mass-radius data are the only ones on the quadratic DEF model with massive scalars~\cite{Tuna:2022qqr}.{Even though we are not aware of any similar studies for the linear DEF model or the sGB model, future studies might change this, which can further help our analysis here.}


{ We mentioned that our conclusions for massless scalars also have implications for low enough scalar masses $m_\phi$. The exact values for which a constraint exists can only be known after a quantitative study, but we expect our results to be relevant for $m_\phi \lesssim 10^{-11}$eV based on the findings for the quadratic DEF model~\cite{Tuna:2022qqr}. That is, our rough bounds on $\betagb$ and $\sigma$ are very likely valid for light scalars as well.}


The two models we considered by no means cover the whole range of scalarization phenomena, of which new examples are discovered almost monthly ~\cite{Doneva:2022yqu,Doneva:2022yqu,Chiba:2021rqa}. However, the linear DEF model and the sGB model are both widely studied, the latter being the most studied topic of scalarization in the last few years. Hence, their comparison to the mass-radius data provides important insight to scalarization overall. 

Our results also provide an informed opinion about other scalarization models we have not yet studied quantitatively. It has been long known that a massive scalar generically suppresses scalarization compared to a massless one, suppression growing with increasing mass~\cite{Ramazanoglu:2016kul}. Thus, we would still not be able to constrain the linear DEF model if the scalar was massive. The sGB model can likely be constrained for small scalar masses as we discussed, but there will be large enough mass values where the constraints would be lost, which is known to be the case for the quadratic DEF model~\cite{Tuna:2022qqr}. In another example, the well-known \emph{asymmetron} model of scalarization uses $A(\phi) = (1-\Delta) +  \Delta\ e^{\beta \phi^2/(2\Delta)}$ with $0 < \Delta \lesssim 1$ in order to explain certain cosmological observations~\cite{Chen:2015zmx}. This form of the coupling function $A(\phi)$ is known to decrease the overall deviations from GR compared to the quadratic DEF model for the same value of $\beta$. Hence, it is almost certain that the constraints on the asymmetron from the mass-radius data would be weaker than those of the quadratic DEF model, analogous to the sGB model. Though, determining if there are constraints at all would require a quantitative study.

What is the applicability, or limitations, of our methodology beyond scalarization? As we discussed, substantial deviations from GR around neutron stars are a prerequisite for the current data to be useful. At first look, the best candidates to repeat our analysis on are theories with analogs of spontaneous scalarization in other fields, such as spontaneous vectorization and tensorization~\cite{Ramazanoglu:2017xbl,Ramazanoglu:2017yun}. However, recent work suggests the uniqueness of scalarization in that tachyonic instabilities of other fields such as vectors imply ill-posed time evolution, rendering any such theories unphysical~\cite{Garcia-Saenz:2021uyv,Silva:2021jya,Demirboga:2021nrc,Doneva:2022ewd}. Hence, these ideas can already be ruled out. Nevertheless, there are various alternative gravity theories which are of interest both in scalar-tensor theories and beyond~\cite{Fujii:2003pa,Langlois:2015cwa,deRham:2014zqa}.

{Lastly, we saw that adding a few stars to our dataset can sharpen our bounds considerably. When a newly obtained mass-radius distribution is distinct from the known ones, we expect its effect on the analysis results to be more significant compared to the case where the distribution closely resembles an already existing one in our dataset. This is the case for PSR J0030+0451 and PSR J0740+6620, as we saw that the constraint on the sGB theory became less significant without them in Sec.~\ref{sec:dataset_size}. In the extreme, even a single novel measurement that is ``orthogonal'' to the current data can be the difference between ruling out a theory or not. This might be a guiding principle for observational studies.}

We explored how to test deviations from GR using neutron star mass-radius measurements, which is an underutilized dataset in our opinion. We hope to investigate even more generic alternative theories of gravity in the future by increasing the precision of our methods as well as incorporating other observational signals.

\acknowledgments
We thank Semih Tuna for his help throughout this project, Daniela Doneva for answering our questions about the neutron star solutions in the sGB model, {and Cole Miller, Anna Watts, Tolga Talha Yıldız for their guidance in the use of the mass-radius data}. We also thank Andrew Coates and K{\i}van\c{c} \"Unl\"ut\"urk for suggestions on the manuscript, which led to substantial improvements. E.S.D. and F.M.R. are supported by Grant No. 122F097 of the Scientific and Technological Research Council of Turkey (T\"{U}B\.{I}TAK).

\appendix
\section{Quadratic DEF model with $\beta>0$}
\label{sec:appendix}
We mentioned that there are scalarized solutions of the $\beta>0$ quadratic DEF model, Eq.~\eqref{eq:A_quadratic}, but they are all thought to be unstable. To the best of our knowledge, the instability of these stars was first discussed in \textcite{Mendes:2014ufa}. However, the unstable stars presented therein all belong to the so-called higher harmonic branches where the scalar field has nodes in its radial profile. Such stars were already known to be unstable even for the quadratic DEF model with $\beta<0$, hence this is not especially surprising in itself. However, later linearized analyses in \textcite{Mendes:2018qwo} strongly indicate that all such scalarized solutions are indeed unstable (especially see the supplemental material Fig.~S1 and the related text).

\begin{figure}[h]
\includegraphics[width=\columnwidth]{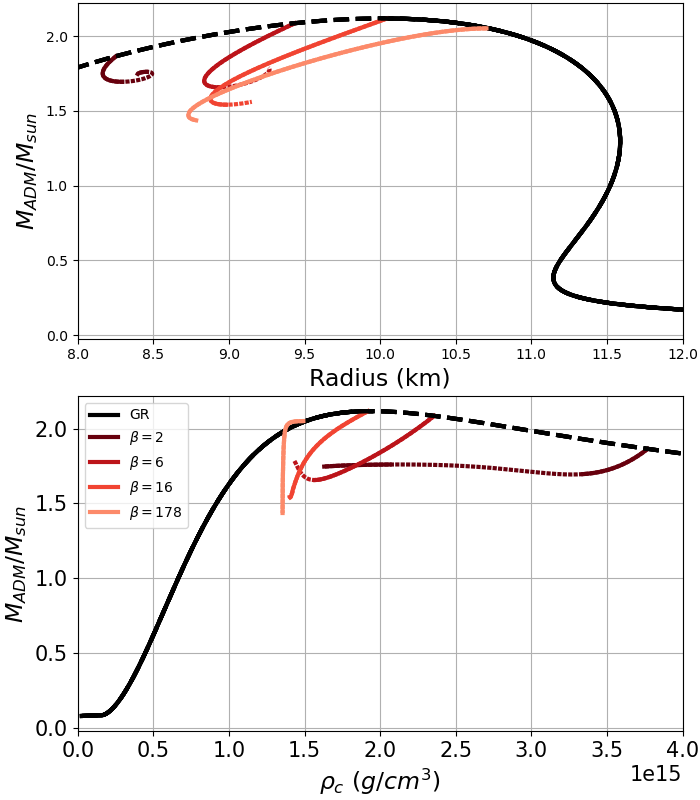}
\caption{Mass-radius and mass-central density diagrams in the quadratic DEF model for various $\beta$. Even though all scalarized solutions are most likely unstable, we only plot the parts with $dM/d\tilde{\rho}_c <0$ with dotted lines.}
\label{fig:quadpos_mr}
\end{figure}

The source of this instability is subtler than the violation of $dM/d\tilde{\rho}_c >0$  in Eq.~\eqref{eq:dmdrho}, which is necessary but not sufficient for stability. We have some of the mass-radius diagrams for the quadratic DEF model with $\beta>0$ in Fig.~\ref{fig:quadpos_mr}, where we see that Eq.~\eqref{eq:dmdrho} is mostly satisfied for scalarized stars. However, studying the linearized oscillations around such solutions shows the existence of exponentially growing modes, which is a direct indication of instability \cite{Mendes:2018qwo}. Hence, these stars are not expected to play an astrophysical role and it is mute to compare them against the mass-radius data. 

\begin{figure}[h]
\includegraphics[width=\columnwidth]{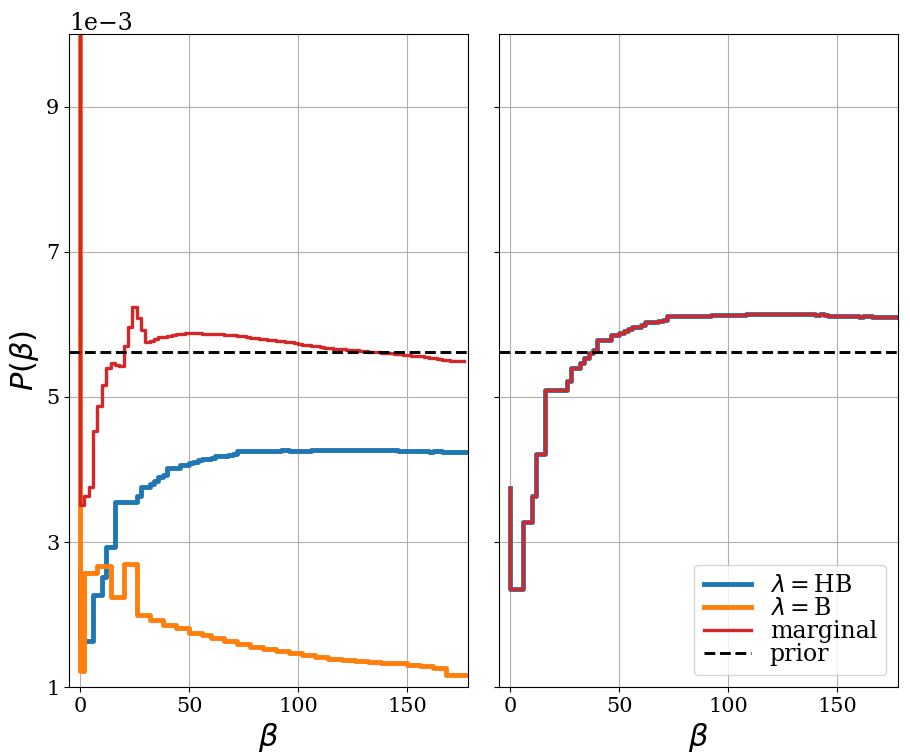}
\caption{Conditional posterior probability for HB, B, and 2B EOS in the quadratic DEF model similar to Fig.~\ref{fig:linearDEFposterior}. Left: the simple flat prior. 2B EOS has a curve very similar to B EOS, but smaller by a factor of $10^{4}$. Right: restricted prior where the only nonzero contribution is due to HB EOS as none of the neutron stars in B and 2B EOS reaches a mass of $2M_{\odot}$.}
\label{fig:quadraticDEF}
\end{figure}

Nevertheless, the fate of these solutions is not sealed until a detailed nonlinear analysis is completed. For this reason, and for the sake of completeness, we repeated the Bayesian analysis for this case by assuming that all scalarized stars satisfying Eq.~\eqref{eq:dmdrho} are stable. We reiterate that this last assumption is most likely wrong. The posterior distributions can be seen in Fig.~\ref{fig:quadraticDEF}. Overall, there is no upper bound on $\beta$ due to a lack of decay in the marginal probability distribution, though there are deviations from the prior at low $\beta$ values. 

\pagebreak

\bibliography{references_all}

\end{document}